\documentclass[pre,preprint,floatfix,amsfonts,showpacs]{revtex4}
\usepackage{graphicx,graphics,color,epsfig}% Include figure files
\usepackage{bm}
\usepackage{amsmath}
\usepackage{amssymb}

\begin{document}

\title{On the Universality of the Energy Response Function in
the Long-Range Spin Glass Model with Sparse, Modular Couplings}

\author{Jeong-Man Park$^{1,2}$ and Michael W. Deem$^1$}

\affiliation{\hbox{}$^1$Department of Physics \& Astronomy\\
Rice University, Houston, TX 77005--1892, USA\\
\hbox{}$^2$Department of Physics, The Catholic University of Korea, Bucheon
420-743, Korea
}

\begin{abstract}
We consider energy relaxation of the long-range
spin glass model with sparse couplings, the so-called dilute 
Sherrington-Kirkpatrick (SK) model, starting
from a random initial state.
We consider the effect that modularity of
the coupling matrix has on this relaxation dynamics.
In the absence of finite size effects, the
relaxation dynamics appears independent of modularity.
For finite sizes, a more modular system
reaches a less favorable energy at long times.
For small system sizes, a more modular system also has a less
favorable energy at short times.
For large system sizes, modularity  may lead
to slightly more favorable
energies at intermediate times.
We discuss these results in the context of evolutionary theory,
where horizontal gene transfer, absent in the Glauber
equilibration dynamics of the SK model studied here, endows
modular organisms with larger response functions
at short times.
\end{abstract}


\maketitle

\section{Introduction}

We here consider energy relaxation in a dilute, modular spin glass.
The form of the energy function, its sparseness and modularity, is
motivated by fitness functions encountered in biology
\cite{Sun2,Dirk,Lipson,Alon2013}.
We emphasize that our calculation is one of statistical mechanics,
rather than of a detailed evolutionary model.
The model is similar in spirit to the spin-glass models that 
have been introduced to analyze the relation between genotype and phenotype
evolution  \cite{Anderson,Kauffman,Kaneko2009,Kaneko2012}.
The model is also quite similar to a model of  associate memory recall,
in which modularity was shown to increase the rate of pattern
matching \cite{Sinha}.
Multi-body contributions to the fitness function in biology, leading
to a rugged fitness landscape and glassy evolutionary dynamics,
are increasingly thought to be an important factor in evolution
\cite{Breen2012}.  That is, biological fitness functions may be
characterized as instances of fitness functions taken from a spin
glass ensemble.  Importantly, though, biological fitness functions
have a modular structure, and their dependence on the underlying
variables is somewhat separable \cite{Bogarad,Deem2005,Sun2006}.
Glassy evolutionary dynamics has been noted a number of times
\cite{Sear,Goldenfeld2006}.
The generalized NK model used to understand the immune response
to vaccines and evolving viruses is a type of modular, dilute
spin glass model \cite{Lee,Park2004,Sun2005,Zhou,Gupta,Yang,Pan}.

We here analyze, within the context of statistical mechanics rather than a
detailed evolutionary model, the dependence of a spin glass response function
on the modularity of the interactions.  
We consider how the spin glass equilibrates from
an initially random state by Glauber dynamics.
At long times, the finite-size corrections to the energy per spin
in the SK spin glass scale as $L^{-2/3}$, where $L$ is the
system size \cite{Parisi1,Parisi2,Billoire,Moore}.  
The timescale for convergence, $t_{\rm ERG}$ grows exponentially
with system size, 
$t_{\rm ERG} \sim t_0 \exp(c L^{1/3})$
\cite{Bouchaud,Horner,Janke,Dotsenko}.

Here, we derive the approximate response function at short times.
Since modularity is a relevant, emergent order
parameter in dynamical systems
\cite{Waddington,Simon,Hartwell1999,Alon,Wagner,Sun}, 
we consider the ensemble of spin glass
Hamiltonians parametrized by modularity, $M$.
In particular, we make predictions for how the energy
relaxation of a dilute spin glass depends on the modularity
of the coupling matrix.
Numerical calculations
have shown that the energy per spin relaxes at different rates
for spin glass systems of different sizes
 \cite{Kinzel}, and these simulations provide additional
motivation for the present calculations.

In a replica calculation, we will show that the response function at
short times is independent of modularity for large system sizes.
This calculation generalizes
the dynamical equations of magnetization and energy
\cite{Sherrington1} 
to the dilute SK model and determines the form that
these equations take near the spin glass phase transition.
The universality of the response function
may be broken by finite size effects.
At long times, greater modularity leads to less
favorable energies due to these finite size effects.
Near the spin glass
transition, there are two opposing finite size effects, and 
greater modularity may lead to a slightly more
rapid energy decay.

The rest of the paper is organized as follows.
In Section \ref{sec2} we describe simple scaling arguments
for the energy relaxation curve
 at short and long
times as a function of modularity.
In Section \ref{sec3}  we introduce the dilute, modular SK model and
the projection of the energy dynamics onto the slow modes.
In Section \ref{sec4}  we derive the slow mode dynamics by a replica approach.
In Section \ref{sec5}  we analyze these equations to produce the energy
relaxation curve.
In Section \ref{sec6}  we use known thermodynamic finite scaling results 
to argue how the dynamical equations depend on system size.
In Section \ref{sec6a} we compare the results to numerical calculations.
We discuss these results in Section \ref{sec7}  and conclude in Section 
\ref{sec8}.

\section{Modularity as a Finite Size Effect}
\label{sec2}
We consider a spin glass with long range couplings.
The entries in the $ N \times N$ coupling matrix
are symmetrically distributed around zero, and the
sum of the variances of the couplings in each row is
$O(1)$.  We contrast this case where every entry of the
matrix may be nonzero to the case where only the entries
along the $L \times L$ block diagonals may be nonzero.  This
latter case is an example of a modular coupling matrix.
The parameter $L$ is a measure of the
effective modularity in the system, with smaller $L$ indicating
greater effective modularity.  

A system with smaller $L$ has
a less favorable ground state energy.
In particular, if we set the negative of the energy per spin to be
$r$, it is known that
$r^* = r_\infty - a L^{-2/3}$ \cite{Moore}.  
The value of
$K$ in the Parisi hierarchy required to stabilize
a system of size $L$ grows as $K \sim (T_c - T) L^{1/6}$, where 
$T$ is temperature \cite{Moore}.  This result
can be used to estimate finite effects if observables are known as
a function of $K$.
In our case,
arguing that the barriers to equilibration of a larger system
further down in the Parisi hierarchy
are of the same order as the energy
of the smaller system from the $K \to \infty$ ground state,
$\Delta E \sim N (r_\infty -r^*)$, we would expect
$t_{\rm ERG} \sim t_0 \exp(c L^{1/3})$
\cite{Bouchaud,Horner,Janke}.
%}, a result with previous theoretical \cite{Horner}
%and numerical support \cite{Janke}.  
We expect logarithmic convergence
to the ground state at long time \cite{Dotsenko}.
Smoothing
the short time behavior, the scaled energy might follow
%\begin{equation}
$r_L(t) = r_\infty - 
a L^{-2/3} \tanh t - 
b [1 + \ln (1 + t/t_{\rm ERG})]^{-2/\nu}
$,
%\label{0a}
%\end{equation}
where $\nu = 1$
to have the expected $L$ dependence at large time, and $a$ and $b$ are
constants of order unity.
%Figure \ref{fig} shows the crossing behavior and illustrates the optimal
%system size as a function of time.  Numerical simulations exhibit
%the energy relaxation as a function of time and system size
%that is shown in this figure \cite{Kinzel}.
The long time ordering of these curves with $L$ is a result of
equilibrium finite size effects.  Whether the curves cross at
short time depends on the details of the equilibration dynamics
and is the subject of the rest of this paper.
%%
%%
%\begin{figure}[t]
%\begin{center}
%\epsfig{file=fig_new.eps,width=0.90\columnwidth,clip=}
%\end{center}
%\caption{Shown 
%is the energy per spin in an equilibrating spin glass 
%of varying size, $L$ (dotted=$5^3$, short dashed=$6^3$, long dashed=$7^3$, and
%solid=$8^3$), as a function of time From Eq.\ (\ref{0a}).
%The system size with the lowest energy 
%is a monotonically increasing function of time.
%}
%\label{fig}
%\end{figure}
%%
%%

%For dynamics such as those in Fig.\ \ref{fig}, there is an optimal  
%system size at a given timescale.
%At short times,
%the system with lowest energy has small $L$ because
%it can reconfigure more quickly.  At 
%long times, the system with the lowest energy has large $L$, because
%more of the phase space is accessible to the connection matrix.
%At intermediate times, the optimal system will
%have an intermediate $L$, with the
%optimal $L$ monotonically increasing with time.

The rest of this article
will calculate the short time behavior of the
energy relaxation curve for a class of coupling matrices
that interpolate between the fully connected $N \times N$ matrix
and one with $L \times L$ block diagonals.  The modularity order parameter,
$M$, is zero in the first case and unity in the  second.

\section{Model}
\label{sec3}
\begin{figure}[t!]
\begin{center}
\includegraphics[width=1in,clip=]{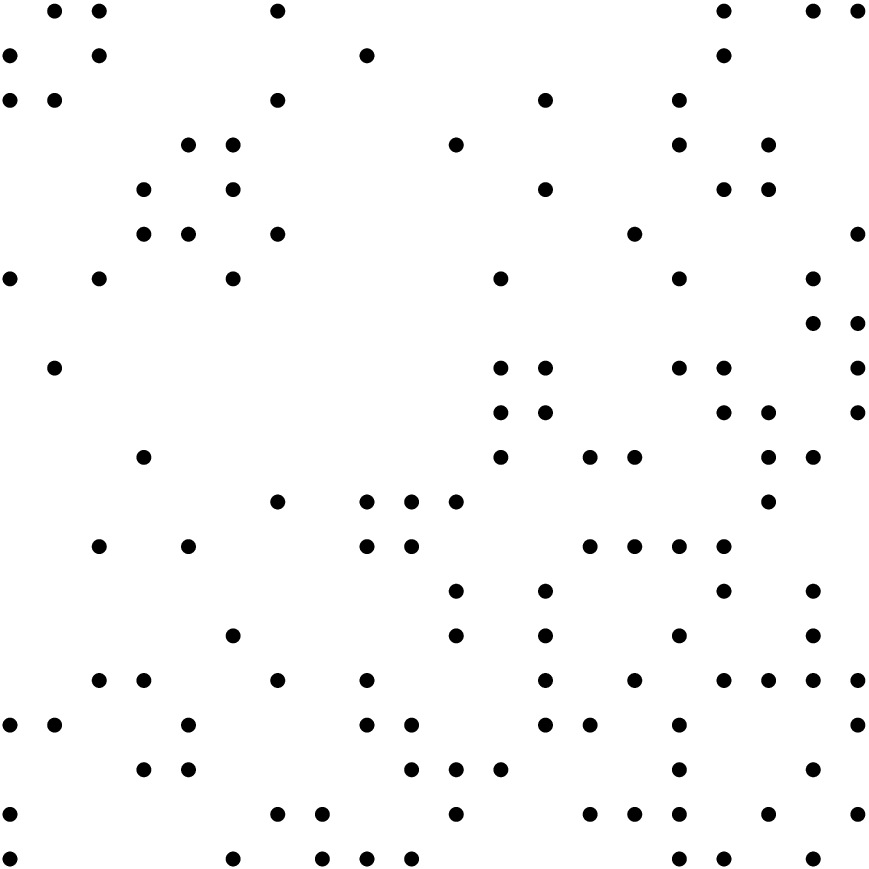}
\hfill
\includegraphics[width=1in,clip=]{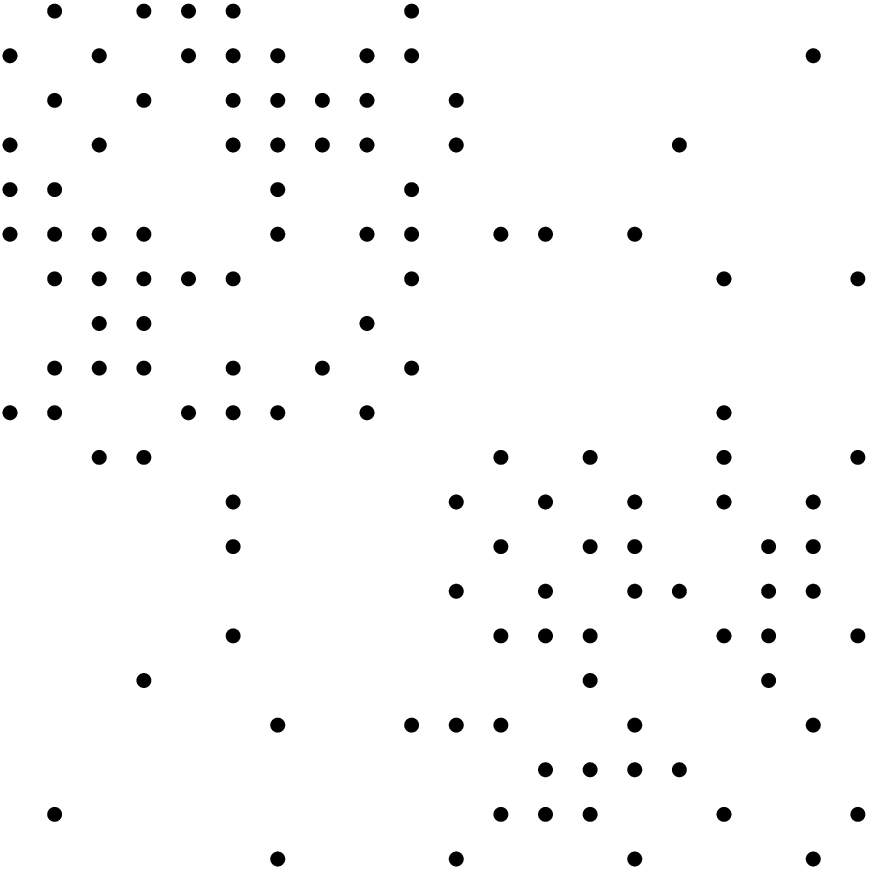}
\hfill
\includegraphics[width=1in,clip=]{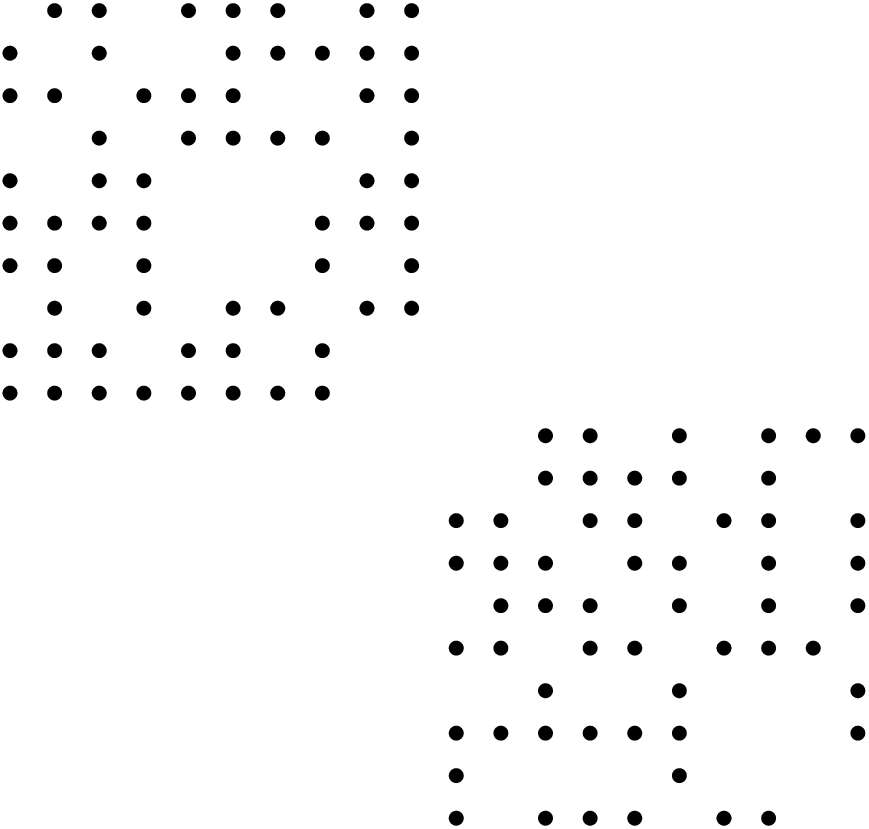}
\end{center}
\caption{Shown is a simplified view of the couplings in
the dilute SK model.  In this figure, we consider a system of
size $N=20$. 
If spin $i$ interacts with spin $j$, 
a dot is displayed at matrix position $i,j$.
Each position $i$ interacts on
average with $C$ other positions.  Here $C = 6$.
Left) A non-modular structure, $M=0$.
Middle) A moderately modular structure, $M=2/3$.
Right) A fully modular structure, $M=1$.
The matrix shown here is the connection matrix, denoted by the symbol $\Delta$.
Here, there are two modules, each of size $L=10$.
We define
modularity from   the excess
number of interactions within the two $L \times L$ block diagonals over that
expected based upon the probability observed outside the block diagonals.
This number is divided by the total number of interactions
to give the modularity, $M$.
}
\label{fig0}
\end{figure}

The focus of the present study is how to introduce modularity
to the SK model, and the resulting short-time dynamics.
The coupling matrix must have local structure, and it must
be sparse, as modularity can not be
identified in a fully connected matrix.
A visual depiction of the non-zero entries in coupling matrix is
shown in Fig.\ \ref{fig0}.

We define a spin glass model that generically incorporates
sparseness and modularity.
The connection matrix for a given system $\alpha$ is denoted
by $\Delta^\alpha$ with $\Delta^\alpha_{ij} = 0,1$,
as shown in Fig.\ \ref{fig0}.
%If we were performing a fully evolutionary calculation we might
%call this system a short protein, with a fold structure indexed
%by $\alpha$ \cite{Sun}.
Each spin $i$ is connected to $C$ other spins, on average.
Putting these points together, 
our simplified model is a dilute SK model:
\begin{equation}
H^\alpha(\{ \sigma \} ) = - \sum_{i<j} J_{ij} \sigma_i \sigma_j
\Delta_{ij}^\alpha
\label{1}
\end{equation}
with $J_{ij} = J z_{ij}$ where z is a quenched Gaussian with zero mean and
variance $1/C$.  The number $C$ is the average number of connections,
and so  in the absence of modularity
$P(\Delta_{ij}) = 
(1-C/N) \delta_{\Delta_{ij}, 0} + (C/N) \delta_{\Delta_{ij}, 1}$.
We have 
$\sigma_i = \pm 1$.
The spin dynamics is governed by Glauber dynamics
such that the rate to flip spin $k$ in the sequence is given by
$w_k( \{ \sigma \}) = \frac{1}{2} ( 1 - \sigma_k 
\tanh \beta h_k )$
where
$h_k = \sum_{j \ne k} J_{kj} \Delta _{kj} \sigma_j = J z_k$,
with
$z_k = \sum_{j \ne k} z_{kj} \Delta_{kj} \sigma_j$.

Now we generalize this model
by introducing modularity, such that
there is an excess of interactions in $\Delta$ along the
$L \times L$ block diagonals of the $N \times N$
connection matrix. There are $k_1 = N/L$ of these
block diagonals.  Thus, 
the probability of a connection is
$C_0/N$ when 
$ \lfloor i/L \rfloor \ne \lfloor j/L \rfloor$ 
and $C_1/N$ when 
$ \lfloor i/L \rfloor = \lfloor  j/L \rfloor$.  The number of 
connections is $C = C_0   + (C_1 - C_0) /k_1$.
Modularity is defined by $M = (C_1 - C_0)  / (k_1 C)$.
To see the spin glass phase, the system must be macroscopic, $N \to \infty$.
In addition, the module size must be large, so that the glass phase appears.
We also require $C$ is large so that the spin glass remains mean field. 

We define the total magnetization 
$m = (1/N) \sum_{i=1}^N \sigma_i$ and
scaled energy per spin
$r = - H / (J N)$.
We split the energy per spin into a component inside
the block diagonals and a component outside:
$r_I = - \sum_{i<j, 
\lfloor i/L \rfloor = \lfloor j/L \rfloor} J_{ij} \sigma_i \sigma_j
\Delta_{ij}^\alpha$, and
$r_O = - \sum_{i<j, 
\lfloor i/L \rfloor \ne \lfloor j/L \rfloor} J_{ij} \sigma_i \sigma_j
\Delta_{ij}^\alpha$, 
with $r = r_I + r_O$.
We also define
$z_k^I = \sum_{j \ne k, \lfloor j/L \rfloor = \lfloor k/L \rfloor}
 z_{kj} \Delta_{kj} \sigma_j$ and
$z_k^O = \sum_{j \ne k, \lfloor j/L \rfloor \ne \lfloor k/L \rfloor}
 z_{kj} \Delta_{kj} \sigma_j$.
We project the microscopic probability of a given state,
$P_t(\sigma)$, onto these order parameters.
These order parameters evolve according to
 \cite{Sherrington1} (see Eqs.\ 8 and 9 therein)
\begin{eqnarray}
\frac{d m } {d t} &=%
\int dx dy D_{m,r_I, r_O;t} [x,y] \tanh \beta J (x+y) - m
\nonumber \\
\frac{d r_I } {d t} &=%
\int dx dy D_{m,r_I, r_O;t} [x,y] x \tanh \beta J (x+y) - 2 r_I
\nonumber \\
\frac{d r_O } {d t} &=%
\int dx dy D_{m,r_I, r_O;t} [z,y] y \tanh \beta J (x+y) - 2 r_O
\label{2}
\end{eqnarray} 
where
\begin{eqnarray}
D_{m,r_I, r_O;t} [x,y] &=&
\lim_{N \to \infty}
\sum_\sigma P_t(\sigma)
\delta([m - m(\sigma)] 
\delta[r_I - r_I(\sigma)]
\delta[r_O - r_O(\sigma)]
\nonumber \\ && \times
\frac
{
\frac{1}{N} \sum_{k=1}^N \delta[x - z_k^I(\sigma)] \delta[y - z_k^O(\sigma)]
}{
\sum_{\sigma'} P_t(\sigma')
\delta([m - m(\sigma')]
\delta[r_I - r_I(\sigma')]
\delta[r_O - r_O(\sigma')]
}
\label{3a}
\end{eqnarray}
We assume that $D_{m,r_I, r_O;t} [x,y]$ is
self-averaging over the disorder, which
numerical simulations out to intermediate times seem to 
support \cite{Sherrington1}.
We will also assume 
equipartitioning
of probability in the macroscopic subshell $(m,r_I, r_O)$
\cite{Sherrington1}.
These assumptions allow us to drop $P_t(\sigma)$
and to perform the averages over the quenched random $z_{ij}$ 
and $\Delta_{ij}$ variables:
\begin{eqnarray}
D_{m,r_I, r_O;t} [x,y] &=&
\lim_{N \to \infty}
\bigg\langle
\sum_\sigma \delta[m - m(\sigma)] 
\delta[r_I - r_I(\sigma)]
\delta[r_O - r_O(\sigma)]
\nonumber \\ && \times
\frac
{
\frac{1}{N} \sum_{k=1}^N \delta[x - z_k^I(\sigma)] \delta[y - z_k^O(\sigma)]
}{
\sum_{\sigma'} \delta[m - m(\sigma')] 
\delta[r_I - r_I(\sigma')]
\delta[r_O - r_O(\sigma')]
}
\bigg\rangle_{ \{ z_{ij} \}, \{ \Delta_{ij} \} }
\label{3}
\end{eqnarray}

\section{Replica Analysis}
\label{sec4}
We now proceed to analytically calculate the averages required to
determine the solution to Eq.\ (\ref{2}).  
We define $w(\sigma) = 
\delta([m - m(\sigma)] 
\delta[r_I - r_I(\sigma)]
\delta[r_O - r_O(\sigma)]$.
 We use the
replica expression in the form
\begin{eqnarray}
\langle
\Phi (\sigma)
\rangle_w &=& 
\frac{
{\rm Tr}_\sigma w(\sigma) \Phi(\sigma)}{{\rm Tr}_\sigma w(\sigma) }
\nonumber \\
&=&
\frac{
{\rm Tr}_{\sigma^1 \ldots \sigma^n}
 w(\sigma^1) \Phi(\sigma^1) w(\sigma^2) \cdots w(\sigma^n)}
{{\rm Tr}_{\sigma^1 \ldots \sigma^n} w(\sigma^1) \cdots w(\sigma^n) 
}
\nonumber \\
&=&
\frac{
{\rm Tr}_{\sigma^1 \ldots \sigma^n}
 w(\sigma^1) \Phi(\sigma^1) w(\sigma^2) \cdots w(\sigma^n)}
{[{\rm Tr}_{\sigma } w(\sigma) ]^n
}
\nonumber \\
&=&
\lim_{n \to 0}
\frac{
{\rm Tr}_{\sigma^1 \ldots \sigma^n}
 w(\sigma^1) \Phi(\sigma^1) w(\sigma^2) \cdots w(\sigma^n)}
{[{\rm Tr}_{\sigma} w(\sigma) ]^n
}
\nonumber \\
&=&
\lim_{n \to 0}
{\rm Tr}_{\sigma^1 \ldots \sigma^n}
 w(\sigma^1) \Phi(\sigma^1) w(\sigma^2) \cdots w(\sigma^n)
\end{eqnarray}
 to write
\begin{eqnarray}
D_{m,r_I,r_O;t} [x,y] &=&
\lim_{N \to \infty} \lim_{n \to 0}
\frac{1}{N} 
\nonumber \\ &&
 \sum_{k=1}^N 
\left\langle
{\rm Tr}_{\sigma^1 \ldots \sigma^n}
\delta[x - z_k^I(\sigma^1)]
\delta[y - z_k^O(\sigma^1)]
w(\sigma^1) w(\sigma^2) \cdots w(\sigma^n)
\right\rangle_{ \{ z_{ij} \}, \{ \Delta_{ij} \} }
\nonumber \\ 
\label{3b}
\end{eqnarray}
Using the Fourier representation of the delta function, we find
 \cite{Sherrington1}
\begin{eqnarray}
D_{m,r_I, r_O} [x,y] &=& \lim_{N \to \infty} \lim_{n \to 0} \frac{1}{N}
\sum_{k=1}^N
\int \frac{d \xi d \eta}{(2 \pi)^2}
\left[
\prod_{\alpha=1}^n
\frac{N d \tilde m_\alpha}{2 \pi}
\frac{
N d \tilde r_\alpha^I
N d \tilde r_\alpha^O
}{(2 \pi)^2} 
\right] 
e^{i \xi x +  i \eta y}
\nonumber \\ &&
{\rm Tr}_\sigma e^{
i N \sum_\alpha[\tilde m_\alpha (m - m(\sigma)) + 
\tilde r_\alpha^I r_I + \tilde r_\alpha^O r_O ]
}
\nonumber \\
&&\times 
\left\langle
e^{
-i \xi \sum_{j \ne k}^I z_{kj} \sigma_j^1 \Delta_{kj} 
-i \eta \sum_{j \ne k}^O z_{kj} \sigma_j^1 \Delta_{kj} 
- i \sum_\alpha \tilde r_\alpha^I \sum_{i<j}^I z_{ij} \sigma_i^\alpha
\sigma_j^\alpha \Delta_{ij}
- i \sum_\alpha \tilde r_\alpha^O \sum_{i<j}^O z_{ij} \sigma_i^\alpha
\sigma_j^\alpha \Delta_{ij}
}
\right\rangle_{ \{ z_{ij} \}, \{ \Delta_{ij} \} }
\nonumber \\ 
\label{4a}
\end{eqnarray}
where in the limits of the sum we have used the notation 
$I$ for restriction inside the block diagonals and 
$O$ to restriction outside the block diagonals.
We average the quantity in brackets over the $\Delta_{ij}$,
setting $k=1$ by permutation symmetry to find
\begin{eqnarray}
\prod_{j=2}^L
\left[
\left( 1 - \frac{C_1}{N} \right) + \frac{C_1}{N}
e^{-i \xi z_{1j} \sigma_j^1 - 
i \sum_\alpha \tilde r_\alpha^I \sigma_1^\alpha z_{1j} \sigma_j^\alpha}
\right]
\nonumber \\
\prod_{j=L+1}^N
\left[
\left( 1 - \frac{C_0}{N} \right) + \frac{C_0}{N}
e^{-i \eta z_{1j} \sigma_j^1 - 
i \sum_\alpha \tilde r_\alpha^O \sigma_1^\alpha z_{1j} \sigma_j^\alpha}
\right]
\nonumber \\
\prod_{1 < i < j}^I
\left[
\left( 1 - \frac{C_1}{N} \right) + \frac{C_1}{N}
e^{- i \sum_\alpha \tilde r_\alpha^I \sigma_i^\alpha z_{ij} \sigma_j^\alpha}
\right]
\nonumber \\
\prod_{1<i<j}^O
\left[
\left( 1 - \frac{C_0}{N} \right) + \frac{C_0}{N}
e^{- i \sum_\alpha \tilde r_\alpha^O \sigma_i^\alpha z_{ij} \sigma_j^\alpha}
\right]
\label{4b}
\end{eqnarray}
Recognizing that $C_0/N$ and $C_1/N$ are small, so that the above
expression can be written in exponential form, Eq.\ (\ref{4a})
becomes
\begin{eqnarray}
&&D_{m,r_I, r_O} [x,y] = \lim_{N \to \infty} \lim_{n \to 0} \frac{1}{N}
\sum_{k=1}^N
\int \frac{d  \xi d \eta}{(2 \pi)^2}
\left[
\prod_{\alpha=1}^n
\frac{N d \tilde m_\alpha}{2 \pi}
\frac{N d \tilde r_\alpha^I}{2 \pi} 
\frac{N d \tilde r_\alpha^O}{2 \pi} 
\right] 
e^{i \xi x +  i \eta y}
\nonumber \\ &&
{\rm Tr}_
\sigma e^{
i N \sum_\alpha[\tilde m_\alpha (m - m(\sigma)) + 
\tilde r_\alpha^I r_I +
\tilde r_\alpha^O r_O]
}
\nonumber \\
&&
e^{
\frac{C_1}{N} \sum_{i<j}^I
\left(
\left\langle
\exp( -i \sum_\alpha \tilde r_\alpha^I \sigma_i^\alpha z_{ij} \sigma_j^\alpha)
\right\rangle_{ \{ z_{ij} \} }
-1 
\right)
}
\nonumber \\
&&
e^{
\frac{C_0}{N} \sum_{i<j}^O
\left(
\left\langle
\exp( -i \sum_\alpha \tilde r_\alpha^O \sigma_i^\alpha z_{ij} \sigma_j^\alpha)
\right\rangle_{ \{ z_{ij} \} }
-1 
\right)
}
\nonumber \\
&&
e^{
\frac{C_1}{N} \sum_{j=2}^L
\left(
\left\langle
\exp(-i \xi z_{1j} \sigma_j^1 
-i \sum_\alpha \tilde r_\alpha^I \sigma_1^\alpha z_{1j} \sigma_j^\alpha)
\right\rangle_{ \{ z_{ij} \} }
- 
\left\langle
\exp( -i \sum_\alpha \tilde r_\alpha^I \sigma_1^\alpha z_{1j} \sigma_j^\alpha)
\right\rangle_{ \{ z_{ij} \} }
\right)
}
\nonumber \\
&&
e^{
\frac{C_0}{N} \sum_{j=L+1}^N
\left(
\left\langle
\exp(-i \eta z_{1j} \sigma_j^1 
-i \sum_\alpha \tilde r_\alpha^O \sigma_1^\alpha z_{1j} \sigma_j^\alpha)
\right\rangle_{ \{ z_{ij} \} }
- 
\left\langle
\exp( -i \sum_\alpha \tilde r_\alpha^O \sigma_1^\alpha z_{1j} \sigma_j^\alpha)
\right\rangle_{ \{ z_{ij} \} }
\right)
}
\label{4}
\end{eqnarray}

%We calculate these averages using replica symmetry (RS), 1-step replica 
%symmetry breaking (1--RSB), and full replica symmetry breaking (FRSB).
We introduce overlap parameters for the whole matrix
and for the block-diagonal part of the matrix as
\begin{eqnarray}
q_{\alpha \beta}^I(\sigma)  &=& 
\frac{1}{L} \sum_{i=1}^L \sigma_i^\alpha \sigma_i^\beta, 
\nonumber \\
q_{\alpha \beta}^O(\sigma)  &=& 
\frac{1}{N-L} \sum_{i=L+1}^N \sigma_i^\alpha \sigma_i^\beta, 
\label{5}
\end{eqnarray}
The four sums inside the exponential in Eq.\ (\ref{4}) sum to
$N \psi[q(\sigma)] + g[\sigma_1, q(\sigma)]$, so that
\begin{eqnarray}
D_{m,r_I,r_O} [x,y] &=& \lim_{N \to \infty} \lim_{n \to 0} \frac{1}{N}
\sum_{k=1}^N
\int \frac{d  \xi d \eta}{(2 \pi)^2}
\left[
\prod_{\alpha=1}^n
\frac{N d \tilde m_\alpha}{2 \pi}
\frac{N d \tilde r_\alpha^I}{2 \pi} 
\frac{N d \tilde r_\alpha^O}{2 \pi} 
\right] 
\nonumber \\ &&
e^{i \xi x +  i \eta y}
 e^{
i N \sum_\alpha[
\tilde m_\alpha m +
\tilde r_\alpha^I r_I +
\tilde r_\alpha^O r_O]
}
\nonumber \\ &&
{\rm Tr}_\sigma e^{
N \psi[q(\sigma)] + g[\sigma_1, q(\sigma)] 
- i \sum_\alpha \tilde m_\alpha \sigma^\alpha
}
\end{eqnarray}
where
\begin{eqnarray}
\psi[q(\sigma)] &=&
\frac{C_1}{2 k_1} \left[ 
(T_0(\tilde r_I) -1)
+ \sum_{\alpha<\beta} T_2^{\alpha \beta} (\tilde r_I)
\left(
\frac{1}{k_1} q_{\alpha \beta}^I(\sigma)^2 +
\frac{k_1 - 1}{k_1} q_{\alpha \beta}^O(\sigma)^2
\right)
\right]
\nonumber \\ &&
\frac{C_0 (k_1-1)}{2 k_1} \left[ 
(T_0(\tilde r_O) -1)
+ \sum_{\alpha<\beta} T_2^{\alpha \beta} (\tilde r_O)
\left(
\frac{2}{k_1} q_{\alpha \beta}^I(\sigma) q_{\alpha \beta}^O(\sigma) +
\frac{k_1 - 2}{k_1} q_{\alpha \beta}^O(\sigma)^2
\right)
\right]
\nonumber \\ && + \ldots
\label{6}
\end{eqnarray}
and
\begin{eqnarray}
g[\sigma_1, q(\sigma)] &=&
\frac{C_1}{k_1} \bigg[
\left(
ChT_0(\xi,\tilde r_I) - T_0(\tilde r_I)
\right)
+ \sum_{\alpha} 
ShT_1^{\alpha} (\xi,\tilde r_I)\sigma_1^\alpha
q_{1 \alpha}^I(\sigma)
\nonumber \\ &&
+ \sum_{\alpha<\beta} 
\left(
ChT_2^{\alpha \beta} (\xi,\tilde r_I) - T_2^{\alpha \beta}(\tilde r_I)
\right)
\sigma_1^\alpha \sigma_1^\beta
q_{\alpha \beta}^I(\sigma)
\bigg]
\nonumber \\ &&
\frac{C_0(k_1-1)}{k_1} \bigg[
\left(
ChT_0(\eta,\tilde r_O) - T_0(\tilde r_O)
\right)
+ \sum_{\alpha} 
ShT_1^{\alpha} (\eta,\tilde r_O)\sigma_1^\alpha
q_{1 \alpha}^O(\sigma)
\nonumber \\ &&
+ \sum_{\alpha<\beta} 
\left(
ChT_2^{\alpha \beta} (\eta,\tilde r_O) - T_2^{\alpha \beta}(\tilde r_O)
\right)
\sigma_1^\alpha \sigma_1^\beta
q_{\alpha \beta}^O(\sigma)
\bigg]
\label{7}
\end{eqnarray}
where terms higher order in the spin overlaps have been omitted.
Here  $T$, $ChT$, and $ShT$ are combinatorial factors:
\begin{eqnarray}
T_k^{\alpha_1 \alpha_2 \cdots \alpha_k}(\tilde r)
&=& \left\langle
\tanh(-i \tilde r_{\alpha_1} z_{ij} )
\cdots
\tanh(-i \tilde r_{\alpha_k} z_{ij} )
\prod_{w=1}^n \cosh(i \tilde r_w z_{ij} )
\right\rangle_{ \{ z_{ij} \} }
\nonumber \\
Ch T_k^{\alpha_1 \alpha_2 \cdots \alpha_k}(x,\tilde r)
&=& \left\langle
\cosh(i x z_{ij})
\tanh(-i \tilde r_{\alpha_1} z_{ij} )
\cdots
\tanh(-i \tilde r_{\alpha_k} z_{ij} )
\prod_{w=1}^n \cosh(i \tilde r_w z_{ij} )
\right\rangle_{ \{ z_{ij} \} }
\nonumber \\
Sh T_k^{\alpha_1 \alpha_2 \cdots \alpha_k}(x,\tilde r)
&=& \left\langle
\sinh(-i x z_{ij})
\tanh(-i \tilde r_{\alpha_1} z_{ij} )
\cdots
\tanh(-i \tilde r_{\alpha_k} z_{ij} )
\prod_{w=1}^n \cosh(i \tilde r_w z_{ij} )
\right\rangle_{ \{ z_{ij} \} }
\end{eqnarray}
Expanding these
in $\rho$ and $1/C$:
\begin{eqnarray}
T_0 &=& 1 + \sum_{w=1}^n \rho_w^2/(2C) + \sum_w \rho_w^4 / (8 C^2) +
\sum_{w < w'} 3 \rho_w^2 \rho_{w'}^2 / (4 C^2) + \ldots
\nonumber \\ 
T_2^{\alpha \beta} &=& \rho_\alpha \rho_\beta / C 
- \rho_\alpha \rho_\beta / C^2 (\rho_\alpha^2 + \rho_\beta^2) 
+ 3 \rho_\alpha \rho_\beta / (2 C^2) \sum_w \rho_w^2 + \ldots
\nonumber \\ 
T_4^{\alpha \beta \gamma \delta} &=& 
(3 / C^2) \rho_\alpha \rho_\beta \rho_\gamma \rho_\delta + \ldots
\nonumber \\ 
ChT_0 &=& 1 - x^2 / (2 C) + x^4 / (8 C^2) + \ldots
\nonumber \\ 
ChT_2 &=& \rho^2/C - 2 \rho^4 / C^2 - 3 \rho^2 x^2 / (2 C^2) + \ldots
\nonumber \\ 
ChT_4 &=& 3 \rho^4 /C^2 + \ldots
\nonumber \\ 
ShT_1 &=& (- i x) \rho/C - (-i x) \rho^3 / C^2 + (-i x)^3 \rho / (2 C^2)
+ \ldots
\nonumber \\ 
 ShT_3 &=& (-i x) 3 \rho^3 / C^2 + \ldots
\end{eqnarray}

Introducing a Fourier representation for the $q$,
we find a final expression of
\begin{eqnarray}
&&D_{m,r_I, r_O} [x,y] = \lim_{N \to \infty} \lim_{n \to 0}
\int \frac{d \xi d \eta}{(2 \pi)^2}
e^{i \xi x  + i \eta y}  
\prod_{\alpha=1}^n
\frac{N d \tilde m_\alpha}{2 \pi}
\frac{N d \tilde r_\alpha^I}{2 \pi} 
\frac{N d \tilde r_\alpha^O}{2 \pi} 
\nonumber \\ &&
\prod_{\beta=1}^n
\frac{N d \tilde q_{\alpha \beta}^I d q_{\alpha \beta}^I }{2 \pi} 
\frac{N d \tilde q_{\alpha \beta}^O d q_{\alpha \beta}^O}{2 \pi} 
e^{N f}
\langle e^{g( \sigma)} \rangle_{X_I(\sigma) }
\label{8}
\end{eqnarray}
where $g(\sigma) = g[\sigma, q(\sigma) \to q]$ and
\begin{eqnarray}
\langle e^{g( \sigma)} \rangle_{X_{I/O}} =
\frac{
{\rm Tr}_\sigma e^{g(\sigma)} e^{X_{I/O}(\sigma)}
}{ 
{\rm Tr}_\sigma e^{X_{I/O}(\sigma)}
}
\label{9}
\end{eqnarray}
Here
\begin{eqnarray}
X_{I} (\sigma) &=& -i \left[
\sum_\alpha \tilde m_\alpha \sigma^\alpha
+ \sum_{\alpha < \beta}^I  \tilde q_{\alpha \beta}^I
 \sigma^\alpha \sigma^\beta
+ \ldots
 \right]
\nonumber \\ 
X_{O} (\sigma) &=& -i \left[
\sum_\alpha \tilde m_\alpha \sigma^\alpha
+ \sum_{\alpha < \beta}^O  \tilde q_{\alpha \beta}^O
 \sigma^\alpha \sigma^\beta 
+ \ldots
\right]
\label{10}
\end{eqnarray}
and the $f$ from Eq.\ (\ref{8})  is given by
\begin{eqnarray}
f &=&
i \sum_\alpha[\tilde m_\alpha m + 
\tilde r_\alpha^I r_I +
\tilde r_\alpha^O r_O ]
+ i \sum_{\alpha < \beta} \left(
 \frac{1}{k_1} \tilde q_{\alpha \beta}^I q_{\alpha \beta}^I  +
 \frac{k_1-1}{k_1} \tilde q_{\alpha \beta}^O q_{\alpha \beta}^O 
\right)
+ \psi[q(\sigma)]
\nonumber \\
&& + \frac{1}{k_1} \ln {\rm Tr}_\sigma e^{X_{I} (\sigma)}
+ \frac{k_1-1}{k_1} \ln {\rm Tr}_\sigma e^{X_{O} (\sigma)}
\label{11}
\end{eqnarray}
In the large $N$ limit, these integrals reduce to a saddle point
calculation, and for stability we find $\tilde m_\alpha = i \mu_\alpha$,
$\tilde r_\alpha^I = i \rho_\alpha^I$,
and $\tilde r_\alpha^O = i \rho_\alpha^O$.

We find 
\begin{eqnarray}
m &= &
\frac{1}{k_1} \langle \sigma_\alpha \rangle_{X_I}
+ \frac{k_1-1}{k_1} \langle \sigma_\alpha \rangle_{X_O}, 
\nonumber \\
r_I &=& \frac{\partial}{\partial \rho_{\alpha}^I} \psi(\rho_I,\rho_O)
\nonumber \\
r_O &=& \frac{\partial}{\partial \rho_{\alpha}^O} \psi(\rho_I,\rho_O)
\label{12}
\end{eqnarray}
here
$
\psi ( \rho_{I}, \rho_{O} ) = \psi [ q(\sigma) \to q,
\tilde{r}^{I}_{\alpha} \to i \rho^{I}_{\alpha},
\tilde{r}^{O}_{\alpha} \to i \rho^{O}_{\alpha} ]
$,
and the overlap parameters to be the expected multipoint averages:
$q_{\alpha \beta}^I = \langle \sigma_\alpha \sigma_\beta \rangle_{X_I}$
and 
$q_{\alpha \beta}^O = \langle \sigma_\alpha \sigma_\beta \rangle_{X_O}$.
We now consider the zero net magnetization case, $m=0$.  
The saddle point conditions become
\begin{eqnarray}
r_I &=& 
\frac{1}{2 a} \rho_I
 \left[
1 + \sum_{1<\beta} \left(
\frac{1}{k_1} {q_{1 \beta}^I}^2 +
\frac{k_1-1}{k_1} {q_{1 \beta}^O}^2
\right)
\right]
\nonumber \\ &&
+O(\rho^3, 1/C^2)
\nonumber \\ 
r_O &=& 
\frac{1}{2 b} \rho_O
 \left[
1 + \sum_{1<\beta} \left(
\frac{2}{k_1} {q_{1 \beta}^I} {q_{1 \beta}^O} +
\frac{k_1-2}{k_1} {q_{1 \beta}^O}^2
\right)
\right]
\nonumber \\ &&
+O(\rho^3, 1/C^2)
\label{13}
\end{eqnarray}
with 
$X_I(\sigma) =\sum_{\alpha < \beta} \rho^2 Q_{\alpha \beta}^I 
\sigma^\alpha \sigma^\beta$ and
$X_O(\sigma) =\sum_{\alpha < \beta} \rho^2 Q_{\alpha \beta}^O 
\sigma^\alpha \sigma^\beta$ 
where
\begin{eqnarray}
\rho^2 Q_{\alpha \beta}^I 
 &=&
\frac{1}{a} \rho_I^2 q_{\alpha \beta}^I +
\frac{1}{b} \rho_O^2 q_{\alpha \beta}^O
\nonumber \\
\rho^2 Q_{\alpha \beta}^O 
 &=&
\frac{1}{a} \rho_I^2 q_{\alpha \beta}^O +
\frac{1}{b(k_1-1)} \rho_O^2 
\left[ q_{\alpha \beta}^I  +
       (k_1-2) q_{\alpha \beta}^O \right]
\label{13a}
\end{eqnarray}
where $1/a = C_1 / (k_1 C)$ and $1/b = C_0(k_1-1)/(k_1 C) = 1-1/a$.
Note that this equation contains order parameters to all orders.
Near the phase transition, we will keep terms to second order in $\rho$.

\section{Dynamical Analysis}
\label{sec5}
We initiate
the dynamical equations (\ref{2}) 
with a random distribution of spins and watch the relaxation
to equilibrium.  
The relaxation 
undergoes a change when
the paramagnetic phase looses stability to the spin glass phase.
At this point $q_I$ and $q_O$ become non-zero.  This happens
when $r = 1/2$.  We are interested in the regime
$r = 1/2 + \epsilon$.  Since $\epsilon$ is small, and since
we have assumed $D$ is self-averaging, we assume replica 
symmetry holds.  The self-consistent equations for the
order parameters are
\begin{eqnarray}
q_I &=&
\int \frac{d u}{\sqrt{2 \pi}} e^{-u^2/2} \tanh^2 \rho \sqrt{Q_I} u
\nonumber \\
q_O &=&
\int \frac{d u}{\sqrt{2 \pi}} e^{-u^2/2} \tanh^2 \rho \sqrt{Q_O} u
\label{13b}
\end{eqnarray}
To  second order in $\epsilon$ these equations have four solutions.
Appendix A shows that the most stable solutions is $q_I = q_O = 0$ for
$r < 1/2$ and $q_I = q_O = q =(4 r^2 - 1) / (32 r^4) \sim 2 \epsilon$
for $r > 1/2$.  Here $\rho_I = \rho_O = \rho$ plays the role
of a time-dependent inverse temperature.

Appendix B shows that $d r_I/dt$ and $d r_O/dt$ satisfy
the same differential equation.   Since they have the same
initial condition $r_I(0) = r_O(0) = 0$, they are proportional.
In fact, we find $a r_I(t) = b r_O(t) = r(t)$.  This result
is expected since it says the average energy inside (outside) the block
diagonals is proportional to the number of connections inside
(outside).
Appendix B shows
\begin{eqnarray}
\frac{dr}{dt} &=&
-2 r - \frac{1}{ \pi} {\rm Re}
\int_{- \infty}^{\infty}  \frac{d \eta}{\eta} 
\frac{d}{d \eta} \bigg\{
e^{-C + C e^{-\eta^2/(2 C)} + 2 i r \eta e^{-\eta^2/(2 C)} }
\nonumber \\ &&
\left[
1 + 2 i r q^2 \eta e^{-\eta^2/(2 C)}
+ 2 r^2 q^2 \eta^2 e^{-\eta^2/ C}
- 8 i r^3 q^2 \eta (1 - \eta^2/C) e^{-\eta^2/C}
+ O(q^3)
\right]
\bigg\}
\nonumber \\ &\sim&
-2r + \sqrt\frac{2}{\pi} e^{-2 r^2} + 2 r {\rm erf}(\sqrt 2 r)
- q^2 \left[
\sqrt\frac{2}{\pi} 2 r^2 e^{-2 r^2} + 2 r (4 r^2-1) {\rm erf}(\sqrt 2 r)
\right]
{\rm ~as~} C \to \infty
\nonumber \\ 
\label{13c}
\end{eqnarray}

\begin{figure}[t]
\begin{center}
\epsfig{file=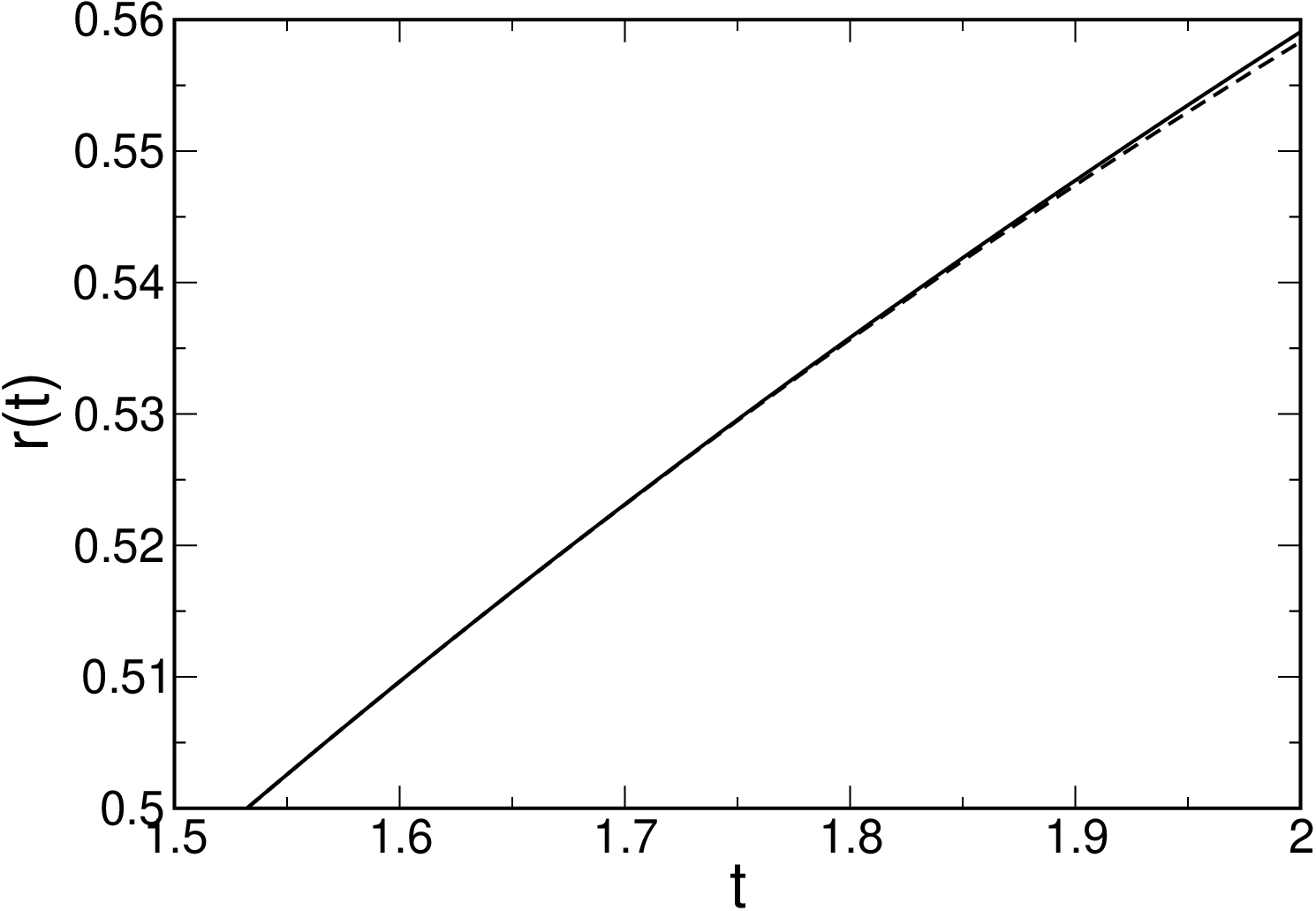,width=0.90\columnwidth,clip=}
\end{center}
\caption{Shown are the paramagnetic (solid, $q=0$) and 
spin glass(dashed, $q>0$) solutions
to Eq.\ (\ref{13c}).
After the critical point at $r=1/2$,
the spin glass phase relaxes more slowly than
does the paramagnetic phase.
Here $C=12$.
}
\label{fig3}
\end{figure}

Figure \ref{fig3} shows how the energy per spin relaxes in the
paramagnetic and spin glass phases.  At $r(t_c) =1/2$, the
spin glass phase emerges.
This occurs at $t_c = \int_0^{1/2} dr/ (dr/dt) \sim 1.439$ as $C \to \infty$.
That is, $r^{\rm SG} = r^{\rm PARA} = 1/2$ at $t = t_c$.
The term proportional to $q^2$ is always negative for $r > 1/2$.
Thus, for $r>1/2$, 
$ r^{\rm SG} <  r^{\rm PARA}$ because
$d r^{\rm SG}/dt <  dr^{\rm PARA}/dt$.
In other words,
the spin glass relaxes more slowly than does the paramagnetic phase
for $t > t_c$.

This calculation suggests that the
energy relaxation is universal, i.e.\ $r(t)$ does not have an
explicit dependence on the modularity, $M$.
Presumably, this is because the effect of modularity is
a finite size effect.
It also happens that projecting the energy onto the $r_I$ and $r_O$
components gives the same result as projecting the energy
onto $r$.  

\section{Finite-Size Corrections to the Dynamics}
\label{sec6}
Finite-size scaling of 
spin glass thermodynamics near the phase transition
has been analyzed by the TAP equations \cite{Bray79}.
The analysis proceeds by analyzing a matrix that at
the transition has the form $A_{ij} = 2 I - J_{ij}$.
The density of eigenvalues, $\lambda$, takes the
form $\rho(\lambda)= \sqrt \lambda / \pi$ for small $\lambda$.
The susceptibility goes as $\chi = \int d \lambda \rho(\lambda) / \lambda^2
\sim 2 \lambda_1^{-1/2} / \pi$ where $\lambda_1$ is the smallest eigenvalue.
It has been argued that finite size
thermodynamics for a spin glass of size $N$ can be 
understood by thermodynamics of an infinite spin glass 
with a finite value of $K$ in the Parisi RSB scheme \cite{Moore}.
It is argued that to stabilize the Gaussian propagator, 
the self-energy in the RSB scheme,
$4 \Delta t^2 / (2K+1)^2/3$,
with $\Delta t = 1 - T/T_c$,
 should be set to the
inverse of the susceptibility, 
calculated above as $\pi \lambda_1^{1/2} / 2$ \cite{Moore,Janis}.
Corrections to the spin coupling parameter
scale as $q = \Delta t+ \Delta t^2 - 2 \Delta t^2 / (2 K+1)^2/3$
\cite{Janis}.
Combining these results, one finds 
\begin{equation}
q = 2 \epsilon - \pi \lambda_1^{1/2} / 4
\end{equation}
The factor $\pi/4$ is only an estimate and may be replaced
by another constant.
For a Gaussian coupling matrix, $\lambda_1 \sim N^{-2/3}$ \cite{Bray79},
and $\lambda_1$
is distributed according to the Tracy-Widom distribution \cite{TW},
This distribution is universal
for matrices with variances equal to the Gaussian ensemble
and symmetric probability distributions \cite{Soshnikov}.
We, thus, conclude
\begin{eqnarray}
q &=& 2 \epsilon - \Delta q,
\nonumber \\
\Delta q &\approx& \pi N^{-1/3}/4
\label{14b}
\end{eqnarray}

Expression (\ref{14b}) tells us the finite size effects on $dr/dt$
for large $N$ for non-modular matrices,  with
$M=0$.  For a perfectly
modular matrix, $M=1$, we can use this expression with
$N \to L$.  
%The constant will change slightly because in this
%case we seek the largest of $N/L$ eigenvalues distributed according
%to the Tracy-Widom distribution for a matrix of size $L$. 
In Appendix C, we show that $\lambda_1$ increases from
the $M=0$ value to the $M=1$ value.
Thus, $q$ will be somewhat smaller in the
$M=1$ case than in the $M=0$ case.
Near $r=1/2 + \epsilon$ , for $C \to \infty$ and $q = O(\epsilon)$,
 the dynamical equation
(\ref{13c}) takes the form
\begin{eqnarray}
\frac{dr}{dt} &=&
-2 r + 1.167 + 1.365 \epsilon 
+ 0.968 \epsilon^2 
- 0.242 q(M)^2
+ O (\epsilon^3)
\label{13e}
\end{eqnarray}
Since $q$ becomes smaller as $M$ increases from 0 to 1, we see that
$r_M(t) > r_{M=0}(t)$ for $ r > 1/2$.
Thus, this calculation suggests
that modularity increases the rate of relaxation
for $t > t_c$.
Interestingly, 
if $q = 2 \epsilon$, a non-vanishing $q$ exactly cancels the 
$O (\epsilon^2)$ term in the
above expression.

\section{Numerical Results}
\label{sec6a}
\begin{figure}[t]
\begin{center}
\epsfig{file=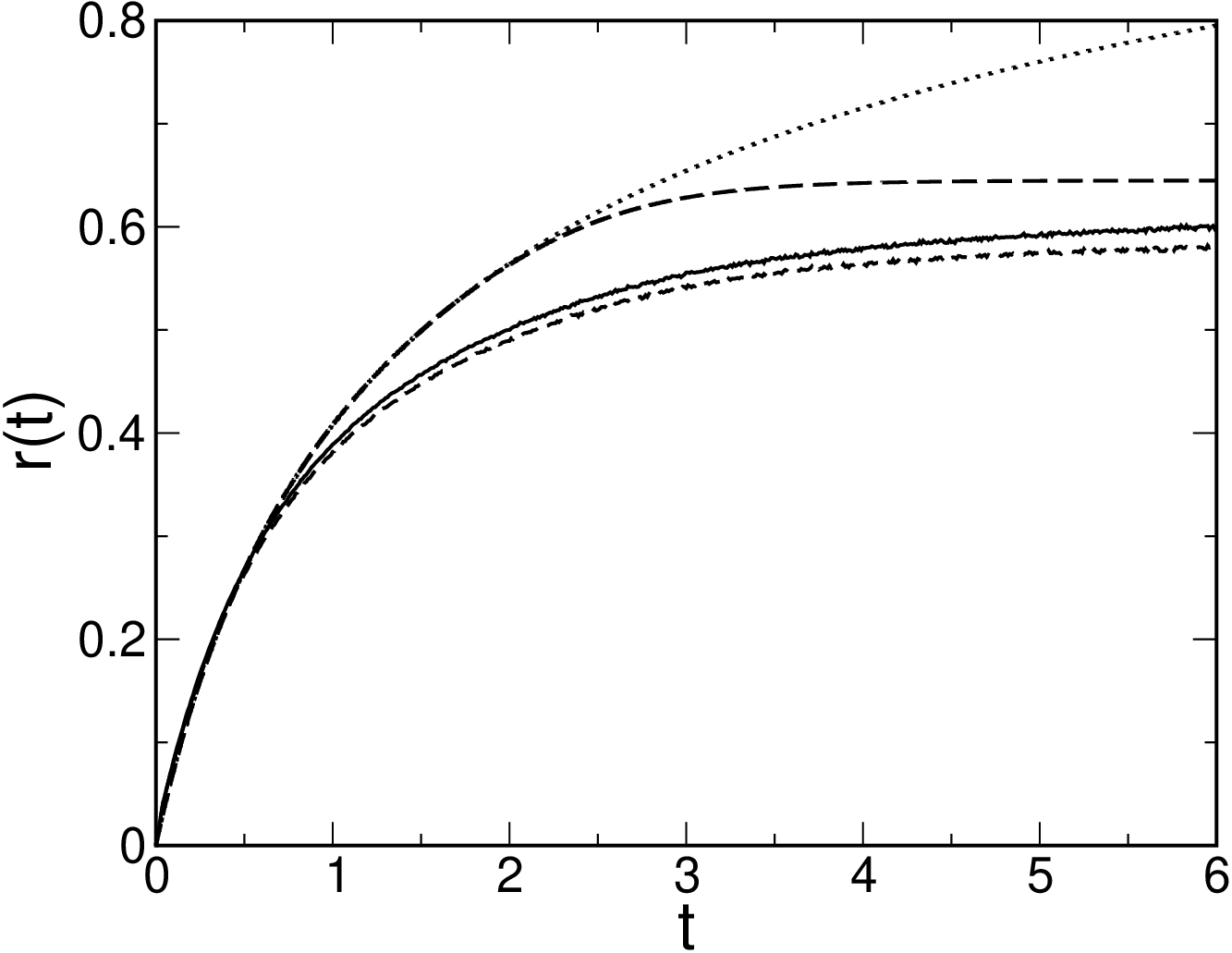,width=0.90\columnwidth,clip=}
\end{center}
\caption{Shown is the $r(t)$ curve for
$N=64, N/L = 4, C=16$ for $M=0$ (solid) and
$M=1$ (short dashed).  
Also shown 
is the prediction of Eq.\ (\ref{13c}) 
for $r^{\rm PARA}(t)$ (dotted) and
$r^{\rm SG}(t)$ (long dashed).
\label{fig7}
}
\end{figure}

We here use a Lebowitz-Gilespie algorithm to sample the
continuous-time Markov process that describes
the Glauber dynamics that lead to Eq.\ \ref{13c}
\cite{Lebowitz75,Gillespie76,Sherrington1}.  We first consider the
case of a small matrix, $N = 64$, with $k_1 = 4, C=16$.
We performed $10^6$ samplings of the Markov process,
collecting the continuous time $r(t)$ curves into
bins in time.  
For large matrices and short times, $t \ll t_c$, the
results reproduce those of Eq.\ (\ref{13c}),  which
are independent of $M$,  in
agreement with previous calculations for $M=0$ \cite{Sherrington2}.
The average results for small matrices
with $M=0$ and $M=1$ are
shown in Fig.\ \ref{fig7}.
The response function of the modular matrix
is below that of the non-modular matrix.

\begin{figure}[t]
\begin{center}
a)
\epsfig{file=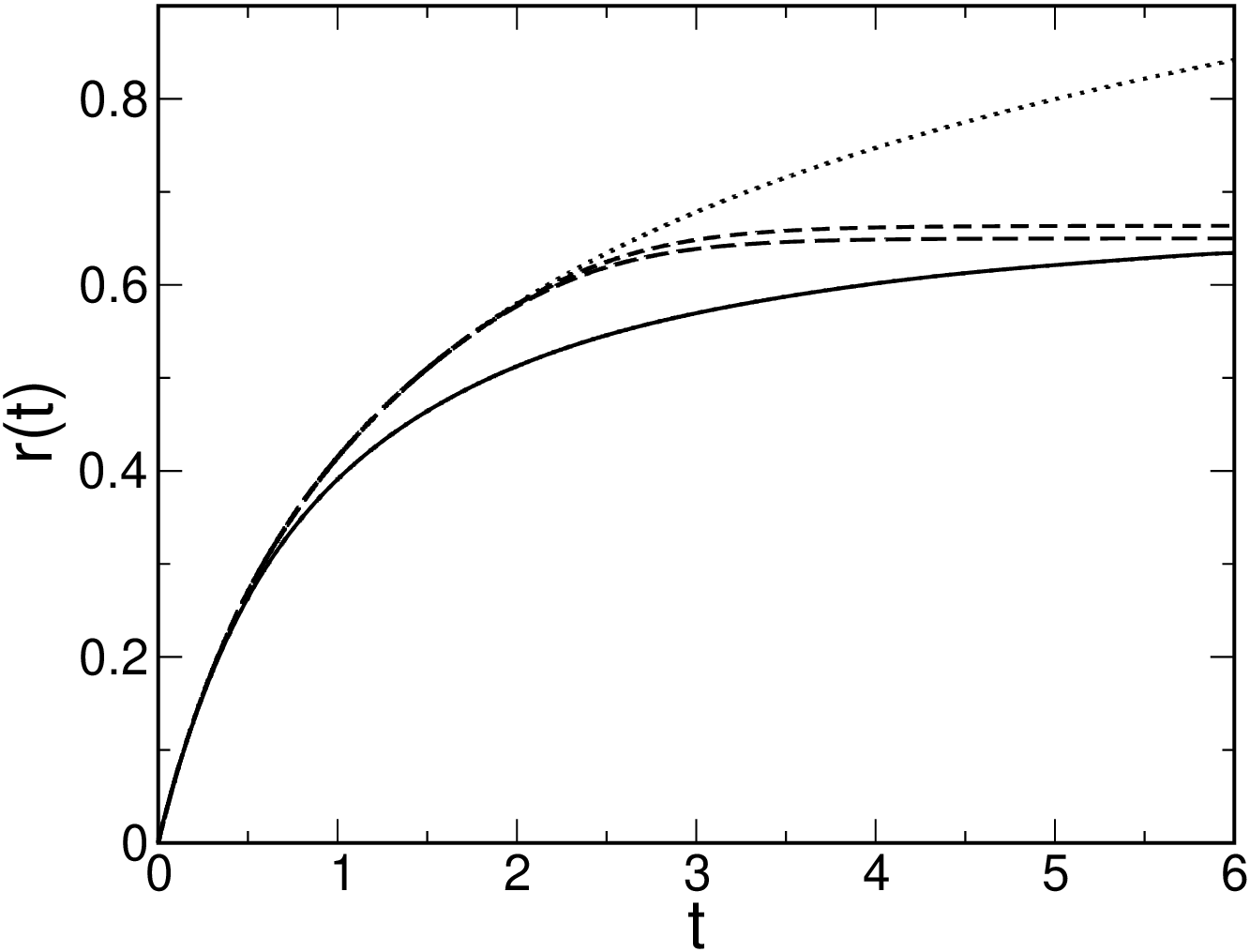,width=0.45\columnwidth,clip=}
b)
\epsfig{file=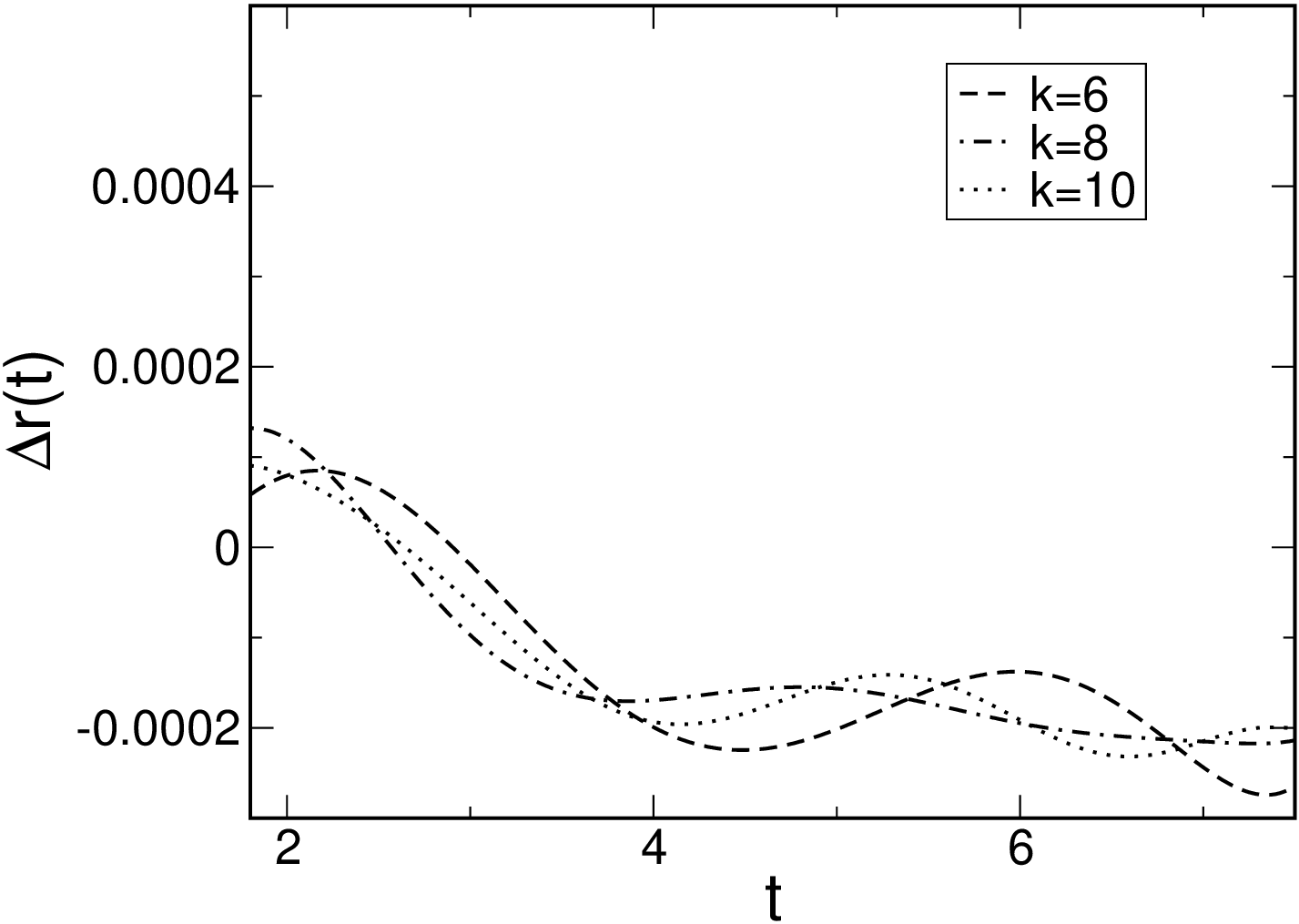,width=0.45\columnwidth,clip=}

\end{center}
\caption{a) 
The $8^{\rm th}$ order polynomial fits
to $r(t)$ for 
$N=16000, N/L = 4, C=4000$ for $M=0, 1$.
The $M=0$ and $M=1$ curves (solid) are indistinguishable on 
this scale.  
Also shown 
is the prediction of Eq.\ (\ref{13c}) 
for $r^{\rm PARA}(t)$ (dotted),
$r^{\rm SG}(t)$ (long dashed), and
$r^{\rm SG}(t)$ using Eq.\ (\ref{14b}) (short dashed).
b) Shown is the 
$k^{\rm th}$ order polynomials curve fits for $k=6,8,10$
to the difference,
$r_{M=1}(t) - r_{M=0}(t)$,
 between 2000 samples of the Markov process.
\label{fig8}
}
\end{figure}

We next consider the case of a large matrix,
$N = 16000$ with $k_1 = 4, C=4000$.  This is a large matrix,
so we performed
$10^3$ samplings of the Markov process
for $M=0$ and $M=1$.
We performed the calculation
independently two times, and the
results are qualitatively similar, with a crossing of the
average $r_{M=1}(t)$ and $r_{M=0}(t)$ curves at some $t > t_c$.
We fit difference between
the continuous time $r(t)$ curves for $t>t_c$ to
$k^{\rm th}$ order polynomials
in time,
shown in Fig.\ \ref{fig8}.
There is an interval after the critical point, 
 $t_c < t < t^* $ during which
the response function of the modular matrix
appears to be above that of the non-modular matrix.
%The standard deviation from the fitted curve
%at any point in time is approximately $0.003$.
%The coefficients of the $k^{\rm th}$ order
%polynomial were fit to approximately $4400$
%data points in time, so the standard error of the curve over
%a timescale 0.5 is roughly $0.00015$.
%The standard error of the maximum value of the $10^{\rm th}$ order
%fit to $\Delta r$ is 0.00024.  
The standard error of the average of the histogrammed points in this range
is $5.1 \times 10^{-5}$.
Thus,
the observed difference between the $M=1$ and $M=0$ response
functions is about two standard errors.
From equilibrium finite size effects, we know $r_{M=1}(t) - r_{M=0}(t)=
\Delta r(t) < 0$ for
large enough $t$, and Fig.\ \ref{fig8}b reproduces this
expected trend.

The projection of the dynamics to $r_I, r_O, m$ in 
Eq.\ (\ref{2}) is approximate.  A more
accurate approximation is obtained by
projecting to the distribution
of local fields \cite{Sherrington2}.
The result is qualitatively similar to Fig.\ \ref{fig8}a:
the spin glass phase emerges at $t_c$ when $q>0$, and
$r^{\rm SG}(t) < r^{\rm PARA}(t)$.  Quantitatively, $t_c$ shifts from
1.439 for $C \to \infty$ to a value $1.85$, also observed in the
numerical simulations here.
We expect that the argument of
Eq.\ (\ref{14b}) will also apply to this more involved calculation,
which again, does not take into account the $t \to \infty$ finite
size effects. We expect that the qualitative conclusions for such
a calculation will be similar to those of Section \ref{sec6}.

\section{Discussion}
\label{sec7}
For Glauber dynamics, the effect of modularity on
the dynamics at short time is a small finite-size effect.
From Figure \ref{fig3} we see that the difference between
the paramagnetic and spin glass dynamics is not large
near $t_c$,
and the effects of modularity are only a small
perturbation of the spin glass dynamics, Eq.\ (\ref{14b}).
At long time, there is a clear effect of modularity,
because the less modular matrix converges to a more stable
energy per spin than does a more modular matrix.
Figure \ref{fig8}b suggests a modest crossing of the $r_{M=1}(t)$ and
$r_{M=0}(t)$ curves after $t_c$.

The results of Fig.\ \ref{fig8} are not dramatic
%the results of Fig.\ \ref{fig} suggested could be possible
and are smaller than 
Eqs.\ (\ref{13c}) and (\ref{14b}) would predict.
Eq.\ (\ref{14b}) is approximate and cannot be used near $r = 1/2$, but if
it is, it predicts an effect $10 \times$ larger than what is
observed in Fig.\ \ref{fig8} at $t = t_c + 0.4$.
What Eqs.\ (\ref{13c}) and (\ref{14b})  miss is the equilibrium
finite size effects for large $t$.  These effects are opposite
in sign to what Eq.\ (\ref{14b}) suggests and
cause $r_M(t)<r_{M=0}(t)$ for large enough $t$.

%Imagine, now, there is selection for the spin glass system to have
%a large response function 
%at timescale $t$, i.e.\ that there is selection for $r(t)$ to be large.
%Further imagine that modularity, $M$, is the variable upon which this
%selection occurs.
%For $t>t_c$, but not too large, this selection will identify
%$M=1$ as optimal, because $r^{\rm SG}_M(t) >
%r^{\rm SG}_{M=0}(t)$.  
%Thus, selection for systems with large short-time
%response function will identify modular systems.
%Conversely, for $t$ very large, 
%this selection will identify $M=0$ as optimal, because 
%$r^{\rm SG}_M(t\infty) \sim r_\infty - a L^{-2/3}$
%whereas
%$r^{\rm SG}_{M=0}(t\infty) \sim r_\infty - a N^{-2/3}$.
%It is natural to expect that at intermediate timescales
%an intermediate value of $M$ will be identified.

In biology horizontal gene transfer significantly enhances
the emergence of modularity 
in different individuals evolving on a common, rugged fitness landscape
\cite{Sun}.  In the spin glass
language, the simple mechanistic
picture is that
different instances of the dynamical
ensemble can find  states that approximately optimize $r$
within one of the $L \times L$ block diagonals.  Horizontal gene transfer
can then combine $N/L$ of these partial solutions of length $L$ into a
near optimal state of length $N$.  This recombination of partial
states is thought to exponentially speed up identification of
optimal states.  Due to the mean field nature of model (\ref{1}),
nucleation of correlations corresponding to ground states in the
modules is averaged out.  Perhaps more significantly, the Glauber dynamics
studied here does not have the multi-spin flip analog
of the horizontal gene transfer move.

\section{Conclusion}
\label{sec8}
We have performed a replica calculation for the
dynamics of a dilute, modular SK model.  Correlations in this model were defined
by a connection matrix, which was parametrized by its modularity.
These calculations suggest
that the energy relaxation of the dilute SK model is 
universal, independent
of the value of modularity for infinite systems.
Finite size arguments show that a
non-modular matrix relaxes to a more
stable energy at long times.
Finite size arguments suggest that the energy relaxation may be
quicker for a modular connection matrix,
possibly leading to more slightly favorable energy
values at intermediate times  near the spin glass transition.
The effect for Glauber dynamics is quite modest.

%We suggested that if the dilute SK model is interpreted as a rough
%model for evolution of biological structure, the present results
%illustrate a selective pressure for modularity to arise in biological
%populations evolving in changing environments,
%in which the system is continuously relaxing from
%new initial conditions.  This interpretation
%rationalizes a number of empirical
%observations for increased modularity in changing environments
%\cite{Dirk}.

Interestingly, in biology horizontal gene transfer significantly enhances
the emergence of modularity, and modularity can
enhance biological fitness \cite{Sun}. 
In the absence of horizontal gene transfer,
modularity does not significantly change fitness
 in these models.
The present statistical mechanics calculations,
showing little dynamic effect of modularity, are
consistent with the latter biological results.
The Glauber dynamics used here do not
contain a multi-spin move that is analogous to 
horizontal gene transfer.
Calculation of the effect of horizontal gene transfer 
for finite, modular biological systems
is an open problem.

\section{Appendix A: Stability of the Overlap function}
We expand Eq.\ (\ref{13b}) to second order in $q_I$ and $q_O$, using
Eq.\ (\ref{13}) and replica symmetry.
  This coupled set of equations can be solved
by the quartic formula to yield four solutions, with
lengthy explicit expressions.  The first
solution is $q_I^A = q_O^A = 0$.  The second solution can be found by setting
$q_I^B = q_O^B = q^B$, with solution $q^B = (4 r^2 -1) / (32 r^4)$.
The third solution can be found by searching for a solution that goes
to zero at $r_0^C$ and is of order $r - r_0^C$.  This yields
an additional solution with $q_I^C \ne q_O^C$ and $r_0^C = 1/(2 \sqrt M)$.
Near $r_0$, this solution looks like
$q_I^C = - (k_1-1) q_O^C = 2 (k_1-1)/(k_1-2) \sqrt M (r - r_0^C)$.
There is a fourth solution that changes from complex to real at
$r_0^D = [1 + 2 (1-M) \sqrt{k_1-1} / (k_1 M)]^{1/2}/2$.
Interestingly $q^C$ also turns from complex to real at $r_0^D$, with
$q_I^C(r_0^D) = q_I^D(r_0^D) $ and
$q_O^C(r_0^D) = q_O^D(r_0^D) $.

The solution that is most stable is the one which extremizes
 (which means maximize as $n \to 0$) the dynamical
free energy.  The dynamical free energy is
\begin{eqnarray}
\beta \overline f &=& - \lim_{n \to 0} f^*/n
\nonumber \\ 
&=&
-\ln 2 + 
\frac{\rho_I^2}{4 a} +
\frac{\rho_O^2}{4 b}
- \frac{3}{4}
\frac{\rho_I^2}{a} \left(
\frac{1}{k_1} q_I^2 + \frac{k_1-1}{k_1} q_O^2
\right)
- \frac{3}{4}
\frac{\rho_O^2}{b} \left(
\frac{2}{k_1} q_I q_O + \frac{k_1-2}{k_1} q_O^2
\right)
\nonumber \\ &&
+
\frac{1}{ 4 k_1}
\left(
\frac{\rho_I^2 q_I}{a} +
\frac{\rho_O^2 q_O}{b} 
\right)^2
+
\frac{k_1-1}{ 4 k_1}
\left(
\frac{\rho_I^2 q_O}{a} +
\frac{\rho_O^2 (q_I + (k_1-1) q_O}{k_1 b} 
\right)^2
\label{14a}
\end{eqnarray}

The dynamical free energy can be evaluated for the four solutions.
We consider $g = \beta \bar f + \ln 2$. We find
$g^A = r^2$.  When $q_I = q_O$, we find $g = r^2 + \epsilon q_I^2$,
so that $g^B = r^2 + 4 \epsilon^3$.
At $r_0^D$, we find $g^C = g^D = 
r^2 - 4 \epsilon^3 [k_1^2 - 4 k_1 (\sqrt{k_1-1} -1) - 4] / 
(8 k_1 \sqrt{k_1-1})$.
The term  proportional to $\epsilon^3 $ in $g^C = g^D$ is always negative,
so that solution B is more stable at $r_0^D$.
At $r_0^C$, $g^C = r^2$.  We find that $g^D(r_0^C) =
{r_0^C}^2 + 4 \epsilon^3 [1 - 8/k_1 + 24/k_1^2 - 32/k_1^3 - 16 / k_1^4]$.
For $k_1>1$, solution B is again most stable.
There does not appear to be a crossing of the C,D free energies
with the more stable B free energy.
We, thus, find solution B is most stable for $r > 1/2$.

\section{Appendix B: $dr_I/dt$ and $dr_O/dt$}
At the saddle point, Eq.\ (\ref{8}) becomes
\begin{eqnarray}
D_{m,r_I, r_O} [x,y] = \lim_{N \to \infty} \lim_{n \to 0}
\int \frac{d \xi d \eta}{(2 \pi)^2}
e^{i \xi x  + i \eta y}  
\frac{{\rm Tr}_\sigma e^{g( \sigma) + X_I(\sigma) }}
     {{\rm Tr}_\sigma e^{             X_I(\sigma) }}
\label{21a}
\end{eqnarray}
since $f^* = - \beta n \bar f \to 0$.
Here $g(\sigma) = 
g_1(\xi, \rho_I, q_I)/a + 
g_1(\eta, \rho_O, q_O)/b 
$
where
\begin{eqnarray}
g_1(x, \rho, q) &=&
C( e^{-x^2/(2C)} -1) 
- i x \rho e^{-x^2/(2C)} \sigma_1
- i x \rho e^{-x^2/(2C)} \sum_{1 < \alpha} q_{1 \alpha} \sigma_\alpha
\nonumber \\ &&
+ \rho^2 (1-x^2/C)  e^{-x^2/(2C)} \sum_{\alpha < \beta}
    q_{\alpha \beta} \sigma_\alpha \sigma_\beta
\end{eqnarray}
We also have
\begin{eqnarray}
X_I(\sigma) = \sum_{\alpha < \beta}
\left[
\frac{\rho_I^2 q_{\alpha \beta}^I}{a}
+
\frac{\rho_O^2 q_{\alpha \beta}^O}{b}
\right]
 \sigma_\alpha \sigma_\beta
\end{eqnarray}
Near the spin glass transition, the $q$ are small. Assuming
replica symmetry, we find 
\begin{eqnarray}
{\rm Tr}_\sigma e^{             X_I(\sigma) } =
1 + n(n-1) (\rho_I^2 q^I/a + \rho_O^2 q_O/b)^2/4 
+ O(q^3) \to 1 {\rm ~as~} n \to 0
\end{eqnarray}
We also find
\begin{eqnarray}
{\rm Tr}_\sigma 
e^{g( \sigma) + X_I(\sigma) } &=&
e^{
C( e^{-\xi^2/(2C)} -1) /a +
C( e^{-\eta^2/(2C)} -1) /b}
{\rm Tr}_\sigma 
e^{f_1 \sigma_1 + 
\sum_{1 < \alpha} f_{ \alpha} \sigma_\alpha
+ \sum_{\alpha < \beta}
    F_{\alpha \beta}  \sigma_\alpha \sigma_\beta
}
\end{eqnarray}
where
\begin{eqnarray}
f_1 &=&
- i \xi \rho_I e^{-\xi^2/(2C)} /a 
- i \eta \rho_O e^{-\eta^2/(2C)} /b 
\nonumber \\
f_\alpha &=&
- i \xi \rho_I e^{-\xi^2/(2C)} q_{1 \alpha}^I /a
- i \eta \rho_O e^{-\eta^2/(2C)} q_{1 \alpha}^O /b
\nonumber \\
F_{\alpha \beta} &=&
 \rho_I^2 (1-\xi^2/C)  e^{-\xi^2/(2C)} q_{\alpha \beta}^I/a +
 \rho_O^2 (1-\eta^2/C)  e^{-\eta^2/(2C)} q_{\alpha \beta}^O/b
\end{eqnarray}
Taking the trace over $\sigma_{\alpha > 1}$, we find
\begin{eqnarray}
{\rm Tr}_\sigma 
e^{g( \sigma) + X_I(\sigma) } &=&
e^{
C( e^{-\xi^2/(2C)} -1) /a +
C( e^{-\eta^2/(2C)} -1) /b}
\nonumber \\ && \times {\rm Tr}_{\sigma_1}
e^{f_1 \sigma_1} \left[
1 + \frac{1}{2} (n-1) f^2 + (n-1) f F \sigma_1 + \frac{(n-1)(n-2)}{2} F^2
+ (n-1) F^2
\right]
\nonumber \\ 
&\to&
e^{
C( e^{-\xi^2/(2C)} -1) /a +
C( e^{-\eta^2/(2C)} -1) /b}
\left[ \left(1 - \frac{f^2}{2} \right) \cosh f_1
- f F \sinh f_1 \right]
 {\rm ~as~} n \to 0
\nonumber \\ 
&\equiv& G(\xi, \eta)
\end{eqnarray}

We consider the dynamical equations (\ref{2}) in the
limit $\beta J \to \infty$, so that 
$ \tanh \beta J (x+y) \to {\rm sgn}(x+y)$.
We can integrate out the $x,y$ dependence in 
Eq.\ (\ref{2}) given Eq.\ (\ref{21a}) by using
integration by parts to see
\begin{eqnarray}
\int dx dy 
 e^{i \xi x  + i \eta y}  
x {\rm sgn} (x+y)
= 2 (2 \pi) \frac{\delta'(\xi - \eta)}{ \eta}
\end{eqnarray}
and
\begin{eqnarray}
\int dx dy 
 e^{i \xi x  + i \eta y}  
y {\rm sgn} (x+y)
= 2 (2 \pi) \frac{\delta'(\eta - \xi)}{ \xi}
\end{eqnarray}
Eq.\ (\ref{2}) and integration by parts leads to
\begin{eqnarray}
\frac{d r_I}{dt} &=&
-2 r_I - \frac{1}{ \pi} {\rm Re}
\int_{- \infty}^{\infty}  \frac{d \eta}{\eta} 
 \left[
\frac{d}{d \xi} G(\xi,\eta)
\right]_{\xi=\eta}
\nonumber \\
\frac{d r_O}{dt} &=&
-2 r_O - \frac{1}{ \pi} {\rm Re}
\int_{- \infty}^{\infty}  \frac{d \xi}{\xi} 
\left[
\frac{d}{d \eta} G(\xi,\eta)
\right]_{\eta=\xi}
\end{eqnarray}
Evaluating these derivatives, we find
\begin{eqnarray}
\frac{d r_I}{dt} &=& \frac{1}{a} \frac{d r}{dt}
\nonumber \\
\frac{d r_O}{dt} &=& \frac{1}{b} \frac{d r}{dt}
\end{eqnarray}
where
\begin{eqnarray}
\frac{dr}{dt} &=&
-2 r - \frac{1}{ \pi} {\rm Re}
\int_{- \infty}^{\infty}  \frac{d \eta}{\eta} 
\frac{d}{d \eta} \bigg\{
e^{-C + C e^{-\eta^2/(2 C)} + i \eta (\rho_I/a + \rho_O/b) e^{-\eta^2/(2 C)} }
\nonumber \\ &&
\bigg[ 1 +
\frac{1}{2} \left( 
\frac{\rho_I q_I}{a} +
\frac{\rho_O q_O}{b}
\right)^2 \eta^2 e^{-\eta^2/ C}
\nonumber \\ &&
- i 
\left( 
\frac{\rho_I q_I}{a} +
\frac{\rho_O q_O}{b}
\right)
\left( 
\frac{\rho_I^2 q_I}{a} +
\frac{\rho_O^2 q_O}{b}
\right)
\eta (1 - \eta^2/C) e^{-\eta^2/C}
\bigg]
\bigg\}
\end{eqnarray}
Using Eq.\ (\ref{13}) and replica symmetry leads to Eq.\ (\ref{13c}).

\section{Appendix C: Distribution of smallest eigenvalue for a modular matrix}
\begin{figure}[t]
\begin{center}
\epsfig{file=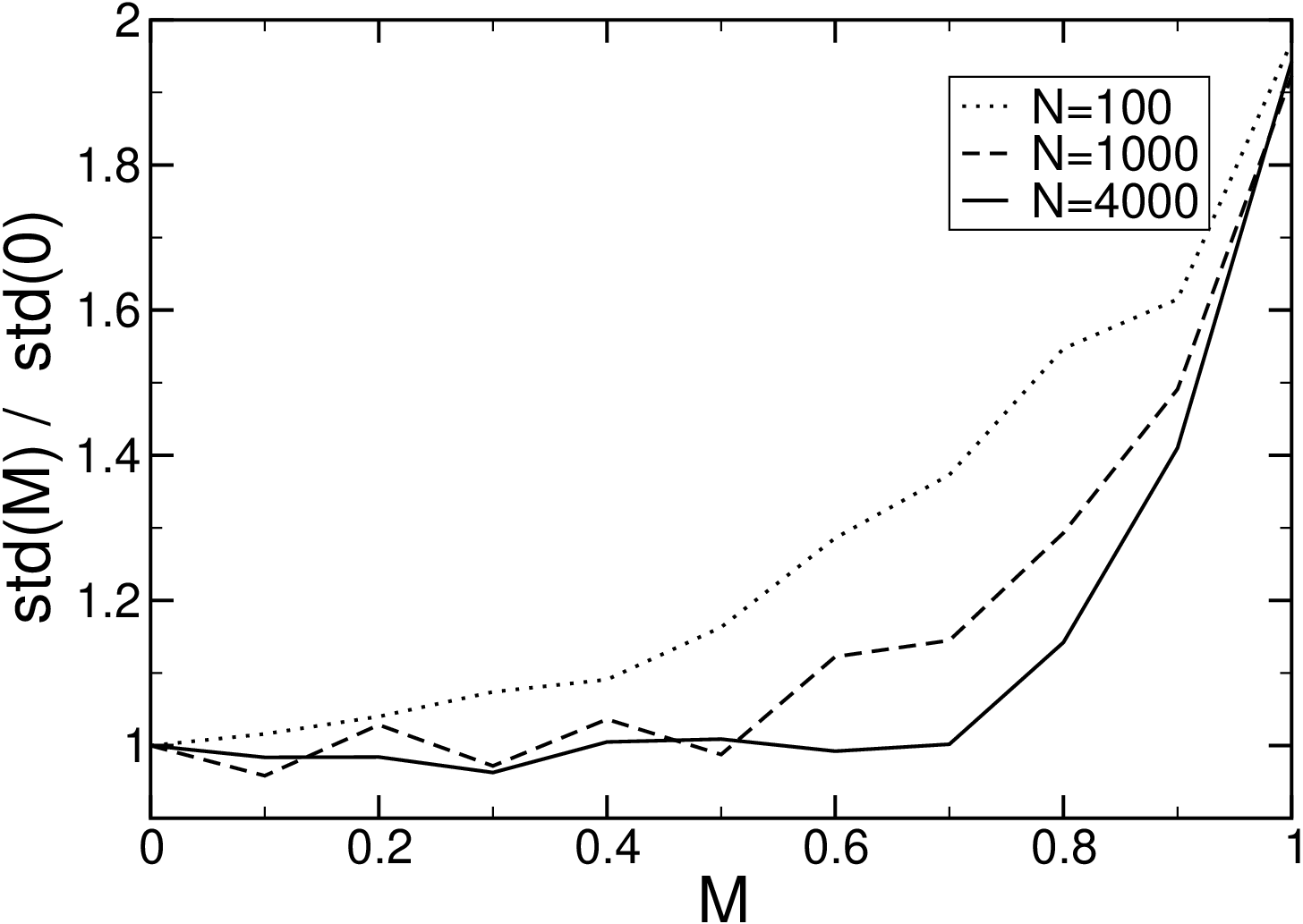,width=0.90\columnwidth,clip=}
\end{center}
\caption{Shown is standard deviation of the smallest eigenvalue of the
matrix defined by Eq.\ (\ref{c1}) as a function of modularity.
Here $N/L = 4$.
\label{fig5}
}
\end{figure}

We here consider how the smallest eigenvalue of a modular
matrix changes from the $N^{-1/3}$ scaling to the
$L^{-1/3}$ scaling as $M$ increases from 0 to 1 in
a random modular matrix.
Up to logarithmic corrections, the density of states and
the distribution of the smallest eigenvalue of any large random
matrix are equivalent to that of a large matrix from the
Gaussian ensemble, essentially as long as $\langle J_{ij} \rangle^2$,
which may depend on $i$ and $j$, 
is the same in the two cases
\cite{Tao2012}.
We, therefore, consider the matrix 
\begin{equation}
B(M) = 2I +\sqrt{M} A_1 + \sqrt{1-M} A_0
\label{c1}
\end{equation}
where $A_1$ is a block diagonal symmetric random Gaussian matrix 
with variance $1/L$ in the $L\times L$ blocks  and
$A_0$ is a $N \times N$ symmetric random Gaussian matrix 
with variance $1/N$ at all entries.  For any $M$, the sum of the variances of in
a row is unity.  We consider the standard deviation of the smallest
eigenvalue of this matrix, $\sigma_M(\lambda_1)$.
We expect $\sigma_M/\sigma_0$ goes from 1 to $d (N/L)^{2/3}$ as
$M$ increases from 0 to 1, where $d$ is the standard deviation of the
maximum of $N/L$ Tracy-Widom random variables divided
by the standard deviation of one Tracy-Widom random variable.
The form of this function is shown in Figure \ref{fig5} for the
case $N/L = 4$.  We see that $\sigma_M/\sigma_0$ increases with $M$.
Whether there is spectral rigidity for $M < M^*$ in the limit $N \to \infty$ 
is unclear \cite{Yau2013}.  Recall that $\sigma_M/\sigma_0>1$ implies the
response curve in Figure \ref{fig3} in the spin glass phase
lies above the curve for $M=0$.

\section*{Acknowledgments} 
This research was partially supported by the US National Institutes
of Health under grant number 1 R01 GM 100468--01 and by
the Catholic University of Korea (Research Fund 2013)
and by the Basic Science Research Program through the National
Research Foundation of Korea (NRF) funded by the
Ministry of Education, Science, and Technology (grant number
2010--0009936).

\bibliography{mod}

\end{document}